\documentclass[12pt]{article}
\usepackage{epsfig}
\usepackage{amssymb}
\usepackage{amsthm}

\theoremstyle{theorem}
\newtheorem{lem}{Lemma}

\newtheorem{prop}[lem]{Proposition}

\theoremstyle{definition}

\newtheorem{pozn}{Remark}

\def\bp{\begin{proof}}
\def\ep{\end{proof}}

\def\be{\begin{equation}}
\def\ee{\end{equation}}
\def\ba{\begin{array}{c}}
\def\ea{\end{array}}

\def\ben{\[}
\def\een{\]}

\newcommand{\bea}{\begin{eqnarray}}
\newcommand{\eea}{\end{eqnarray}}

\newcommand{\kt}{\rangle}
\newcommand{\br}{\langle}
\begin{document}

\begin{center}

{\Large \bf {Solvable model of quantum phase transitions and the
symbolic-manipulation-based study of its multiply degenerate
exceptional points and of their unfolding
 }}

\vspace{10mm}

 {\bf Miloslav Znojil}

 \vspace{3mm}
Nuclear Physics Institute ASCR,

250 68 \v{R}e\v{z}, Czech Republic

{e-mail: znojil@ujf.cas.cz}




\end{center}

\vspace{5mm}

\section*{Abstract}

It is known that the practical use of non-Hermitian (i.e.,
typically, ${\cal PT}-$symmetric) phenomenological quantum
Hamiltonians $H \neq H^\dagger$ requires an efficient reconstruction
of an {\em ad hoc} Hilbert-space metric $\Theta=\Theta(H)$ which
would render the time-evolution unitary. Once one considers just the
$N-$dimensional matrix toy models $H=H^{(N)}$, the matrix elements
of $\Theta(H)$ may be defined via a coupled set of $N^2$ polynomial
equations. Their solution is a typical task for computer-assisted
symbolic manipulations. The feasibility of such a model-completion
construction is illustrated here via a discrete square well model
$H=p^2+V$ endowed with a $k-$parametric close-to-the-boundary
interaction $V$. The model is shown to possess (possibly, multiply
degenerate) exceptional points marking the phase transitions which
are attributable, due to the exact solvability of the model at any
$N<\infty$, to the loss of the regularity of the metric. In the
parameter-dependence of the energy spectrum near these singularities
one encounters a broad variety of alternative, topologically
non-equivalent scenarios.



\section{Introduction}

In the two most recent collections \cite{fort} of papers on the
applicability of non-Hermitian operators in quantum physics one can
find multiple samples of the advantages which are provided by the
use of manifestly non-Hermitian effective quantum Hamiltonians $H
\neq H^\dagger$ in several areas of phenomenology. Among these
advantages one of the key roles is played by the capability of these
sufficiently general phenomenological Hamiltonians of mimicking the
quantum phase transitions and/or an onset of quantum chaos in
many-body systems, etc. The growth of popularity of this area of
research motivated also our present study in which we intend to pay
particular attention to the role of computer-assisted symbolic
manipulations.

From the point of view of mathematics an explanation of the deepest
essence of at least some of the above-mentioned phenomena is not too
difficult since many of them are simply caused by the so called
spontaneous breakdown of certain symmetries. For the most elementary
illustration let us recall, e.g., paper \cite{return} where we
explained that and how the spontaneous breakdown of the combined
parity and time-reversal symmetry (conveniently abbreviated as
${\cal PT}-$symmetry in the physics literature \cite{Carl}) plays
the role of a trigger of transition between the observability and
non-observability of the energy in an elementary toy model of
quantum dynamics. The message delivered by this and similar
elementary examples is nontrivial and unexpectedly deep showing,
e.g., that the possibility of transitions between different
dynamical regimes is closely connected to the presence of
branch-point singularities, say, on the Riemann energy surface
$\mathbb{E}(g)$ in the complex plane of coupling (or of any other
tunable parameter) $g$.

In the closely related Kato's monograph \cite{Kato} on the
mathematics of perturbation theory the latter singularities were
systematically studied via finite-dimensional matrix models and they
were also given the well-chosen name of ``exceptional points'' (EP).
The same finite-dimensional-matrix methodical strategy will be also
accepted in what follows.  In Introduction let us also mention that
in a broader mathematical setting of the geometric singularity
theory one can find the same (or at least very similar) concepts in
the Thom's classical theory of catastrophes \cite{Arnold} (with
multisided applications \cite{Zeeman}) as well as in its multiple
newer descendants: {\em pars pro toto} let us mention our recent
proposals \cite{Chen,catast} of the simplest possible quantum
analogues of such a classical singularity classification pattern.

In the mathematically narrower square-root-branch EP context the
studies of the Riemann-surface singularities found particularly
numerous explicit applications in quantum physics. In the context of
perturbative quantum field theory and in a way enhancing our
understanding of quantum anharmonic oscillations in potentials
$V(x)=x^2+gx^{2+\delta}$ the singularities of this class became
widely known under a nickname of ``Bender-Wu singularities''
\cite{BWTurbiner}. In optics, the alternative theoretical
identification of EPs with the points of non-Hermitian degeneracies
\cite{Berry} encountered an enormous experimental popularity
recently \cite{Makris}. This success was supported not only by the
availability of innovated metamaterials possessing anomalous
refraction indices but also by the underlying analogies between the
Maxwell and Schr\"{o}dinger equations in the dynamical regime of
phase transitions~\cite{Nature}.

After a return to the standard quantum mechanics of stable systems
or to the atomic, molecular or nuclear phenomenology \cite{Nimrod},
the studies of concrete models reveal the existence of EP
hypersurfaces $\partial {\cal D}$ playing the role of certain
natural horizons of observability of quantum systems (i.e., of
certain separation boundaries between different phases), with
numerous important physical as well as mathematical consequences
\cite{Geyer,horizon}. In our present paper we shall be inspired by
this particular problem. We shall describe some of its aspects in
detail, emphasizing that their clarification finds a very natural
methodical support in the symbolic as well as advanced numerical
manipulations mediated, typically, by MAPLE \cite{Maple} and/or by
similar, mostly commercially available software.

\section{The concept of hidden Hermiticity
of Hamiltonians}

During practically all of the history of quantum theory it has been
overlooked that its applicability is restricted by the use of
concrete representations of the physical Hilbert space ${\cal
H}^{(P)}$ {\em made in parallel} with a concrete self-adjoint
representation of observables (say, generators $\mathfrak{h}^{(P)}$
of the unitary evolution with time). A criticism of such a paradigm
emerged, e.g., in Refs.~\cite{Carl,ali}.

The change of the paradigm has been encouraged by the practical
needs of applications of quantum theory in nuclear physics. Besides
the often cited Dyson-inspired non-Hermitian variational approach to
the so called interacting boson models of heavy nuclei \cite{Geyer}
one might also recall another manifestly non-Hermitian variational
method based on the judicious, Hilbert-space-metric-employing
coupling of clusters \cite{Bishop}, etc. One of the main obstacles
of the necessary conceptual separation of the {\em simultaneous}
choices of the Hilbert spaces and Hamiltonian operators was the
difficulty of its implementation in calculations. Only too often,
people prefer the choice of the simplest Hilbert-space
representations (with, say, ${\cal H}^{(P)}\ \equiv \
L^2(\mathbb{R}^d)$) {\em and} of the simplest dynamics (cf. also the
critical commentary in \cite{Hoo} in this context).

On abstract level, the amended quantum theory admitting a broader
class of quantum dynamics may be found summarized in \cite{SIGMA}.
In the spirit of Ref.~\cite{Geyer} one finds the {\em simultaneous}
introduction of space ${\cal H}^{(P)}$ and self-adjoint observable
$\mathfrak{h}^{(P)}$ overrestrictive. The information about dynamics
is separated into the choice of Hilbert space ${\cal H}^{(F)}$
(which remains friendly, cf. the superscript) and a given
Hamiltonian operator $H$ (which need not necessarily remain
self-adjoint in the same space, $H \neq H^\dagger$). The emerging
apparent puzzle (``does one violate the requirements of unitarity
and Stone's theorem?'') has an elementary explanation
\cite{Carl,Geyer}: The initial Hilbert space (say, ${\cal H}^{(F)}\
\equiv \ L^2(\mathbb{R})$) is reclassified as auxiliary and
unphysical. In parallel, the apparently non-Hermitian Hamiltonian
(take, for illustration, just the most popular Bessis' imaginary
cubic oscillator $H^{(ICO)} = p^2+{\rm i}x^3$ \cite{DB} with real
spectrum \cite{DDT}) is also reclassified. As ``crypto-Hermitian'',
i.e., by definition \cite{SIGMA}, as self-adjoint in another,
``standard'' Hilbert space ${\cal H}^{(S)}$.

One makes the Hamiltonian $H$ self-adjoint in ${\cal H}^{(S)}$ by
using just an {\em ad hoc} redefinition of the inner product in
${\cal H}^{(F)}$. Once we have the two different operators $H$ and
$H^\dagger \neq H$ acting on the ket vectors $| \phi \kt$ in ${\cal
H}^{(F)}$, we simply change
 \be
 \br \psi | \phi \kt^{(F)}\ \ \to \ \ \br \psi | \phi \kt^{(S)}:=
 \br \psi |\Theta| \phi \kt^{(F)}\,.
 \label{rapl}
 \ee
The self-adjoint and positive definite operator
$\Theta=\Theta^\dagger>0$ may be perceived as playing the role of
the Hilbert space metric (the mathematical conditions are listed,
say, in \cite{Geyer}). Thus, the usual Hermiticity  of observables
$\Lambda$ is now required in ${\cal H}^{(S)}$,
 \be
 \Lambda^\ddagger := \Theta^{-1}\Lambda^\dagger \Theta = \Lambda\,.
 \label{crypl}
 \ee
For the Hamiltonian $\Lambda_0= H$ (with real spectrum), in
particular, the hidden Hermiticity property
 \be
 H^\dagger \Theta = \Theta\,H\,
 \label{cryplik}
 \ee
is often re-read as an implicit, ambiguous \cite{Geyer} {\em
definition} of a suitable metric $\Theta=\Theta(H)$.

\section{An update of the concept of solvability}


The concept of solvability is often restricted to the availability
of the closed-form eigenstates of ordinary differential Hamiltonians
\cite{Ushveridze}. The phenomenologically oriented search for the
measurable aspects of quantum systems forced many physicists to
search for various extensions of the concept. The question
reemerged, recently, in connection with the new wave of interest in
phase transitions described in terms of the spontaneous breakdown of
antilinear symmetries. Typically, the {\em same} ${\cal
PT}-$symmetric Hamiltonian $H =H(\lambda)$ is considered before and
after the phase transition. During the phase transition itself such
a Hamiltonian becomes ``anomalous'' (i.e., some of its eigenvalues
get complex). In other words, one {\em leaves} the physical domain
of parameters ${\cal D}$.

%
%
%

During the application of such an idea to the above-mentioned
imaginary cubic oscillator $H^{(ICO)}$ people encountered serious
mathematical (i.e., technical \cite{cubic} as well as much more
serious conceptual \cite{Siegl}) difficulties. A partial escape out
of such a trap has been found in the exceptional
differential-operator and boundary-delta-function model
$H^{(Robin)}$ of Refs. \cite{david} which proved sufficiently
representative though still completely solvable \cite{Zelezny}.

A more systematic realization of the project (i.e., in the language
of mathematics, of an exhaustive solution of the operator
Eq.~(\ref{cryplik})) has been revealed and described, in
Refs.~\cite{complete,completeb}, as based on the use of suitable
finite-dimensional toy models.
%
%
Their solvability opened new perspectives in a choice of the
Hilbert-space metric in Eq.~(\ref{crypl}). The key point was that
one did not need to start from a metric which would be given {\em a
priori}. Even in a generalized (e.g., ${\cal PT}-$symmetric) setting
one was suddenly able to reconstruct the metric from the given
Hamiltonian $H \neq H^\dagger$ {\em non-numerically}.

One of the key difficulties now emerged in connection with the
ambiguity of the general solution $\Theta=\Theta_\kappa(H)$ of
Eq.~(\ref{cryplik}). Here the subscripted (multi)index $\kappa$
numbers the alternatives (see \cite{Geyer} for an exhaustive
discussion of this point). The new problem only remains reasonably
tractable in the finite-dimensional Hilbert spaces with $\dim {\cal
H}^{( F,S)}=N<\infty$. In these cases it appears sufficient
\cite{SIGMAdva} to solve the conjugate Schr\"{o}dinger equation
 \be
 H^\dagger |\Xi_n \kt =  E^\dagger_n |\Xi_n \kt
 \ee
(note that by assumption the spectrum is real and discrete here).
One then {\em defines} the general metric by the formula
 \be
 \Theta_{\vec{\kappa}}(H) = \sum_{n=1}^N\, |\Xi_n \kt
  \,\kappa_n \,\br \Xi_n|\,.
  \label{spekex}
 \ee
An {\em arbitrary} optional $N-$plet of coefficients $\kappa_n>0$ is
admitted. In other words, the solvability of the model proves
crucial under the $N<\infty$ assumption.


\subsection{Oscillator-type solvable models}

Even if one decides to work with dimensions $N < \infty$, serious
technical difficulties with the analysis of spectra of $H$ and/or
with the explicit construction of the metrics already emerge at
dimensions as low as $N=4$ \cite{four}. It is recommended to work
with the matrices of specific forms as sampled by the well-motivated
non-Hermitian and PT-symmetric version of the popular Bose-Hubbard
complex Hamiltonian $H^{(BH)}$ \cite{Hubb} and by some other
realistic physical models \cite{Hubbb}.

Alas, strictly speaking, many of the realistic choices of these
computationally tractable (i.e., complex and, typically, tridiagonal
or pentadiagonal) Hamiltonians $H$ cease to be solvable, in spite of
their frequent merit of being well described by perturbation theory
\cite{Hubb}. In this sense the first decisive step toward the
exactly solvable family has only been made in Ref.~\cite{maximal}
where we picked up and studied the anharmonic-oscillator-related
{\em real} and tridiagonal anharmonic-like matrices
  \be
 H^{(N)}_{(ATM)}{}
 =\left [\begin {array}{cccccc}
  1-N&{}\,{g}_1{}&0&0&\ldots&0\\
 -{}\,g_1{}& 3-N&{}\,{g}_{2}{}&0&\ldots&0\\
 0&-{}\,g_{2}{}&5-N&\ddots&\ddots&\vdots
 \\
 0&0&-{}\,{g}_{3}{}&\ddots&{}\,{g}_{2}{}&0
 \\
 \vdots&\vdots&\ddots&\ddots&N-3&{}\,{g}_{1}{}\\
 0&0&\ldots&0&-{}\,g_{1}{}&N-1
 \end {array}\right ]\,
 \label{NyTSpt}
 \ee
which appeared particularly construction-friendly. With the purpose
of their further necessary simplification at the larger $N$ we
imposed an additional requirement ${g}_{N-1}{}={g}_{1}{}$,
${g}_{N-2}{}={g}_{2}{}$, \ldots, having enhanced the symmetry of the
underlying multiparametric coupling pattern perceivably. We were
rewarded by the discovery of the exact solvability of the resulting
model at all $N<\infty$~\cite{maximal}.

This discovery proved heavily dependent on the availability of the
symbolic manipulations with polynomials in MAPLE. {\em Pars pro
toto} let us recall that for the very localization of the degenerate
EP value of the very first coupling ${g}_{N-1}{}={g}_{1}{} :=
\sqrt{D}$ (i.e., of the first coordinate of the vertex of the
boundary manifold $\partial {\cal D}$) at the not too large sample
dimension $N=8$ we had to determine this particular EP value (equal,
incidentally, exactly to $\sqrt{7}$) as a unique ({\em sic}!) root
of the sixteenth-degree secular-like polynomial equation
 \ben
 314432\,D^{17}-5932158016\,D^{16}+
 4574211144896\,{D}^{15}
+3133529909492864\,{D}^{14}+ \een \ben +917318495163561932\,{D}^{13}
+167556261648918275684\,{D}^{12}+ \een \ben
+14670346929744822064505\,{D}^{11}
+720991093724510065469933\,{D}^{10}+ \een \ben
+62429137451114251409236415\,{D}^9
+676326278232758784369966787\,{D}^8+ \een \ben
+40525434802944282153115803370\,{D}^7
+2361976444746440513605248930610\,{D}^6- \een \ben
-145759836636885012145070948315366\,{D}^5+ \een \ben
+8129925258122948689157916436170874\,{D}^4+ \een \ben
-68875673245487669398850290405642067\,{D}^3+ \een \ben
+235326754101824439936800228806905073\,{D}^2- \een \ben
-453762279414621179815552897029039797\,{D}+ \een \ben
+153712881941946532798614648361265167=0\,.
 \een
This and similar polynomials were generated by means of the
Groebner-basis technique as implemented in MAPLE. Needless to add,
even the very proof of the uniqueness of this root (which we never
published, due to its length) required an even more extensive use of
the MAPLE software. Let us also add that the similar
computer-assisted constructions of the EP boundaries had to be
performed just at a few not too large dimensions $N$. Due to the
immanent friendliness of our highly symmetric toy models
$H^{(N)}_{(TAM)}$ we were able to extrapolate the resulting closed
formulae to all $N$, we clarified their structure and we pointed out
their relevance in Ref.~\cite{tridiagonal}. The readers may find
more details therein.

\subsection{Classical-orthogonal-polynomials-related solvable models \label{opole}}

The above family of solvable physics-motivated quantum models
$H^{(N)}_{(TAM)}$ has been complemented by the other, mathematically
motivated and classical-orthogonal-polynomials-related models of
Refs.~\cite{Laguerre}. Subsequently \cite{gegenb} we added more
details of the symbolic-manipulation-assisted constructions of the
necessary Hilbert-space metrics $\Theta$ for the underlying specific
Hamiltonians. For illustration we choose there the
Gegenbauer-related family of  Hamiltonians
 \be
 \left( \begin {array}{cccccc}
 0&1/2\,{a}^{-1}&0&0&\ldots&0
\\
\noalign{\medskip}2\,{\frac {a}{2\,a+2}}&0& \left( 2\,a+2 \right) ^{
-1}&0&\ldots&\vdots
\\
\noalign{\medskip}0&{\frac {2\,a+1}{2\,a+4}}&0& \left( 2 \,a+4
\right) ^{-1}&\ddots&0
\\
\noalign{\medskip}0&0&{\frac {2\,a+2}{2\,a+ 6}}&\ddots& \ddots&0
\\
\noalign{\medskip}\vdots&\ddots&\ddots&\ddots&0& \left( 2\,a+2N-4
\right) ^{-1}
\\
\noalign{\medskip}0&\ldots&0&0&{\frac {2\,a+N-1}{2\,a+2N-2}}&0
\end {array} \right)\,.
 \label{hamil}
 \ee
This enabled us to explain that besides the above-mentioned key
contribution of computer facilities to the feasibility of the
symbolic-manipulation constructions of the eligible metrics
$\Theta$, an equally important role appeared to be played by the
MAPLE-supported numerical software which enables one to control the
numerical precision needed in the, in general, very ill-conditioned
task of the localization of the eigenvalues of $\Theta$. Yielding
the guarantee of the necessary strict positivity of all of these
eigenvalues and of their inverse values. The resulting explicit
knowledge of the boundaries of the domain at which the eigenvalues
of the metric $\Theta=\Theta_\kappa(H)$ were losing their positivity
appeared to be of a particular relevance in the quantum analogue of
the classical theory of catastrophes as described in
Ref.~\cite{catast}.

\subsection{Quantum-graph solvable models}

Just for completeness let us also mention our quantum-graph
proposals \cite{graph} in which the third family of the  solvable
quantum models emerged after a suitable discretization of
coordinates, $x \in \mathbb{R}$ $ \to $ $x_j$, $j \in \mathbb{Z}$.
This trick helped us to make the approximate models tractable by the
standard tools of linear algebra.

The simplest dynamically nontrivial though still topologically
trivial model of the latter discrete-coordinate family dates already
back to Ref.~\cite{complete}. In this paper the most elementary
special case was characterized by the following next-to-trivial
tridiagonal matrix Hamiltonian
 \be
  H^{(N)}({\lambda})=
  \left( \begin {array}{cccccc}
 2&-1-{\it {\lambda}}&0&\ldots&0&0
\\
{}-1+{\it {\lambda}}&2&-1&0&\ldots&0
\\
{}0&-1&\ \ \ 2\ \ \ &\ddots&\ddots&\vdots
\\
{}\vdots&0&\ddots&\ \ \ \ddots\ \ \ &-1&0
\\
{}0&\vdots&\ddots&-1&2&- 1+{\it {\lambda}}
\\
{}0&0&\ldots&0&-1-{\it {\lambda}}&2
\end {array}
 \right )\,.
 \label{toym}
 \ee
In our present paper we intend to generalize such a one-parametric
boundary-interaction family, keeping in mind, i.a., the not yet
explored possibility of connecting this and similar $N<\infty$
quantum Hamiltonians and bound-state spectra with their respective
analogues as defined and derived in continuous limits~\cite{frag}.

\section{New, $k-$parametric boundary-interaction toy model \label{newmodel}}


The highly restricted flexibility of the exactly solvable
one-parametric discrete square well model (\ref{toym}) of
Ref.~\cite{complete} is disappointing. This disappointment is only
partially compensated by the immanent merit of the possible
connection of the model to its continuous $N \to \infty$ limit and
analogue as proposed and described in Refs.~\cite{david}. On the
other hand, one of the key shortcomings of the one-parametric and
discrete $N<\infty$ models (\ref{toym}) is that they do not allow us
to perform any EP-degeneracy fine-tuning, found and accessible in
several multiparametric $N<\infty$ toy models, say, of
Refs.~\cite{Hubb,maximal}. In our recent paper \cite{Wu} we turned
attention, therefore, to the two- and three-parametric extensions of
the above-mentioned model (\ref{toym}). We revealed that the
extended models remain solvable. In our present further extension of
the latter paper we shall introduce and study, therefore, the
entirely general family of the $k-$parametric and $N$ by $N$
dimensional matrix quantum Hamiltonians
%
%
 \be
 H^{(N)}({\lambda, -\mu,\ldots})=\left( \begin {array}{ccccccc}
  2&-1-{\it {\lambda}}&0&\ldots&&\ldots&0
 \\{}-1+{\it {\lambda}}&2&-1+{\it {\mu}}&0&\ldots&&\vdots
 \\{}0&-1-{\it {\mu}}&2&-1-{\nu}&0&\ldots&
 \\{}\vdots&0&-1+{\nu}&2&\ddots&\ddots&\vdots
 \\{}&&\ddots&\ddots&\ddots&-1-\mu&0
  \\{}\vdots&&&\ddots&-1+\mu&2&-1+{\it {\lambda}}
 \\{}0&\ldots&\ldots&0&0&-1-{\it {\lambda}}&2
 \end {array}
 \right)\,.
 \label{hlham}
 \ee
They contain an antisymmetrized and sign-changing sequence of the
couplings $\lambda, \mu,\ldots$ entering the elements
$$-1-{\lambda},-1+{\mu},-1-{\nu},\ldots,  -1+{\nu},-1-{\mu},
-1+{\lambda}$$ of the upper diagonal and, {\em mutatis mutandis},
also the elements of the lower diagonal.

\subsection{The
construction of the simplest Hermitizing metric}

Our first result may be now formulated as the statement that models
(\ref{hlham}) remain solvable at any number of parameters $k$. In
order to illustrate the contents of such a result, let us now take
the sufficiently representative $N=11$ sample of Hamiltonian
(\ref{hlham}) containing, say, the four non-vanishing couplings in
the matrix $H^{(11)}({\lambda, -\mu,\nu,-\rho})$ of the bidiagonal
form with vanishing main diagonal and with the upper diagonal such
that
 $$
 \left \{ 1+ H^{(11)}_{j,j+1}({\lambda, -\mu,\nu,-\rho})\,,\ j=1,2,\ldots,N-1 \right \}
 = \{-\lambda, \mu,-\nu,\rho,0,0,-\rho,\nu,-\mu, \lambda\}
\,
 $$
and with the lower diagonal such that
 $$
 \left \{ 1+ H^{(11)}_{j+1,j}({\lambda, -\mu,\nu,-\rho})\,,\ j=1,2,\ldots,N-1 \right \}
 = \{\lambda, -\mu,\nu,-\rho,0,0,\rho,-\nu,\mu, -\lambda\}
\,.
 $$
We may feel inspired by papers \cite{complete,fortin} and use the
diagonal ansatz for the metric,
 $$
 \left \{\Theta^{(diag)}_{j,j}\,,\ j=1,2,\ldots,N \right \}
 = \{z_1, z_2,z_3,z_4,1,1,\ldots,1,z_4,
 z_3,z_2,z_1\}\,.
 $$
Its insertion converts the crypto-Hermiticity condition
(\ref{cryplik}) into a set of coupled nonlinear algebraic equations.
Their more or less routine solution (using symbolic manipulations)
leads to the unambiguous step-by-step elimination and specification
of the unknown metric-matrix elements,
 $$z_4={\frac {1+\rho}{1-\rho}}\ \equiv \ f(-\rho)\,,\ \ \ \ \ z_3={\frac
{-1+\nu-\rho+\nu\,\rho}{-1+\rho-\nu+\nu\,\rho}}=f(-\rho)\,f(\nu)\,,$$
$$z_2= -{\frac
{-1-\mu+\nu+\nu\,\mu-\rho-\rho\,\mu+\nu\,\rho+\nu\,\rho\,\mu}{
1-\rho+\nu-\nu\,\rho-\mu+\rho\,\mu-\nu\,\mu+\nu\,\rho\,\mu}}
=f(-\rho)\,f(\nu)\,f(-\mu)$$ plus, finally, $z_1= $
$$={\frac
{1-\lambda+\mu-\mu\,\lambda-\nu+\nu\,\lambda-\nu\,\mu+\nu\,\mu
\,\lambda+\rho-\rho\,\lambda+\rho\,\mu-\rho\,\mu\,\lambda-\nu\,\rho+
\nu\,\rho\,\lambda-\nu\,\rho\,\mu+\nu\,\rho\,\mu\,\lambda}{1-\rho+\nu-
\nu\,\rho-\mu+\rho\,\mu-\nu\,\mu+\nu\,\rho\,\mu+\lambda-\rho\,\lambda+
\nu\,\lambda-\nu\,\rho\,\lambda-\mu\,\lambda+\rho\,\mu\,\lambda-\nu\,
\mu\,\lambda+\nu\,\rho\,\mu\,\lambda}}=$$
$$=f(-\rho)\,f(\nu)\,f(-\mu)\,f(\lambda).$$ The extrapolation pattern
is now obvious, yielding the proof of the following, entirely
general

\begin{prop}
Every $k-$parametric Hamiltonian
$H^{(N)}=H^{(N)}(\lambda_1,\lambda_2,\ldots,\lambda_k)$ of
Eq.~(\ref{hlham}) with the real parameters $\lambda_1=+\lambda$,
$\lambda_2=-\mu$, $\lambda_3=+\nu$  etc which are all smaller than
one in absolute value is self-adjoint in Hilbert space ${\cal
H}^{(S)} \sim \mathbb{R}^N$ using the diagonal non-Dirac Hilbert
space metric $\Theta^{(diag)}\neq I$ which differs from the unit
matrix just by its $2k$ outermost diagonal matrix elements
$$\Theta_{kk}^{(diag)}=\Theta_{N+1-k,N+1-k}^{(diag)}=f(\lambda_k),$$
$$
\Theta_{k-1k-1}^{(diag)}=\Theta_{N+2-k,N+2-k}^{(diag)}
=f(\lambda_k)f(\lambda_{k-1}),
$$
$$\ldots,$$
$$\Theta_{11}^{(diag)}=\Theta_{N,N}^{(diag)}
=f(\lambda_k)f(\lambda_{k-1})\ldots f(\lambda_{1}) $$ where
$f(x):=(1-x)/(1+x)$.
\end{prop}

\begin{pozn}
It is more than appropriate to add here that once we managed to
construct the positive and invertible diagonal metric
$\Theta^{(diag)}$, we need not bother about the proof of the reality
of the spectrum of the related crypto-Hermitian matrix
$H^{(N)}=H^{(N)}(\lambda_1,\lambda_2,\ldots,\lambda_k)$ anymore.
Indeed, the latter matrix is, by construction, self-adjoint in the
``new auxiliary'' Hilbert space ${\cal H}^{(NA)} \sim \mathbb{R}^N$
which is assumed endowed with the $^{(S)}-$superscripted inner
product (\ref{rapl}) where $\Theta=\Theta^{(diag)}\neq I$. Following
paper \cite{fortin} one should add that the latter space {\em need
not} coincide with the ultimate physical Hilbert space ${\cal
H}^{(S)}$ of Ref.~\cite{SIGMA}. Still, the (hidden) Hermiticity
$H=H^\ddagger$ in  ${\cal H}^{(NA)}$ {\em implies} the reality of
the energy spectrum of course.
\end{pozn}

\subsection{Non-equivalent
  tridiagonal
 Hermitizing metrics}

In the spirit of paper \cite{complete} let us now recall the highly
ambiguous nature of the general, $N-$parametric,
spectral-expansion-resembling formula (\ref{spekex}) for the metric
and let us make use of this great flexibility in assuming that there
might exist some still sufficiently elementary next-to-diagonal
metric of the form
 \be
 \Theta^{(tridiag)}=\Theta^{(diag)}+v\,{\cal P}
 \label{triton}
 \ee
in which the $N-$dimensional and real diagonal metric
$\Theta^{(diag)}=\Theta^{(diag)}({\lambda_1, \ldots,\lambda_k})$ of
Proposition 1 is ``perturbed'' by a suitable bidiagonal
pseudometric. For the sake of clarity let us first set $N=11$ and
insert the ansatz
 $$
 {\cal P}=
 \left[ \begin {array}{ccccccccccc} 0&{\it t_1}&0&0&0&0&0&0&0&0&0
\\\noalign{\medskip}{\it t_1}&0&{\it t_2}&0&0&0&0&0&0&0&0
\\\noalign{\medskip}0&{\it t_2}&0&{\it t_3}&0&0&0&0&0&0&0
\\\noalign{\medskip}0&0&{\it t_3}&0&t_4&0&0&0&0&0&0
\\\noalign{\medskip}0&0&0&t_4&0&1&0&0&0&0&0\\\noalign{\medskip}0&0&0
&0&1&0&1&0&0&0&0\\\noalign{\medskip}0&0&0&0&0&1&0&t_4&0&0&0
\\\noalign{\medskip}0&0&0&0&0&0&t_4&0&{\it t_3}&0&0
\\\noalign{\medskip}0&0&0&0&0&0&0&{\it t_3}&0&{\it t_2}&0
\\\noalign{\medskip}0&0&0&0&0&0&0&0&{\it t_2}&0&{\it t_1}
\\\noalign{\medskip}0&0&0&0&0&0&0&0&0&{\it t_1}&0\end {array}
\right]\,
 $$
in Eq.~(\ref{cryplik}) again. In a more or less routine manner we
may again solve the resulting set of the $N^2=121$ coupled algebraic
equations yielding the following unique result,
 $$t_4=1+\rho \ \equiv \  (1-\rho)\,z_4\,,\ \ \ \
 t_3={\frac {-1+\nu-\rho+\nu\,\rho}{-1+\rho}} \ \equiv \
(1+\nu)\,z_3\,,$$ $$ t_2={\frac
{-1-\mu+\nu+\nu\,\mu-\rho-\rho\,\mu+\nu\,\rho+\nu\,\rho\,\mu}{-
1+\rho-\nu+\nu\,\rho}} \ \equiv \  (1-\mu)\,z_2$$ plus, similarly
and finally, $t_1= (1+\lambda)\,z_1$. It is rather easy to
generalize now this construction and to reformulate it into a
detailed proof of the following

\begin{prop}
Every $k-$parametric Hamiltonian
$H^{(N)}=H^{(N)}(\lambda_1,\lambda_2,\ldots,\lambda_k)$  is also
self-adjoint in another Hilbert space ${\cal H}^{(S)} \sim
\mathbb{R}^N$ where a tridiagonal non-Dirac Hilbert space metric
$\Theta^{(tridiag)}\neq I$ is used in the form of a positive
definite linear combination (\ref{triton}) of the diagonal metric of
preceding Proposition with the bidiagonal pseudometric ${\cal P}$.
In the latter matrix the main diagonal vanishes while its
non-vanishing upper and lower diagonals have the same form filled
with units, with the exception of the $2k$ outermost matrix elements
$${\cal P}_{kk+1}={\cal P}_{N-k,N+1-k}=(1+\lambda_k)\,f(\lambda_k),$$
$$
{\cal P}_{k-1k}={\cal
P}_{N+1-k,N+2-k}=(1+\lambda_{k-1})\,f(\lambda_k)f(\lambda_{k-1}),
$$
$$\ldots,$$
$${\cal P}_{12}={\cal P}_{N-1,N}=(1+\lambda_{1})\,f(\lambda_k)f(\lambda_{k-1})\ldots
f(\lambda_{1}) $$ where the function $f(x):=(1-x)/(1+x)$ is the same
as above.
\end{prop}

\begin{pozn}
Whenever the real parameter $v$ in formula (\ref{triton}) remains
sufficiently small, the resulting tridiagonal metric
$\Theta^{(tridiag)}$ remains ``acceptable'', i.e., safely positive
and invertible. For the larger values of $v$ the analysis is more
difficult. One must proceed in full methodical parallel with the
analogous problem as studied in Ref.~\cite{gegenb} and mentioned
also in paragraph \ref{opole} above.

\end{pozn}

\section{The descriptive and
spectral properties of the model}

\subsection{The EP degeneracy phenomenon from $H^{(N)}({\lambda, -\mu,\ldots})$ using
$N=11$}

\begin{figure}[h]                     
\begin{center}                         
\epsfig{file=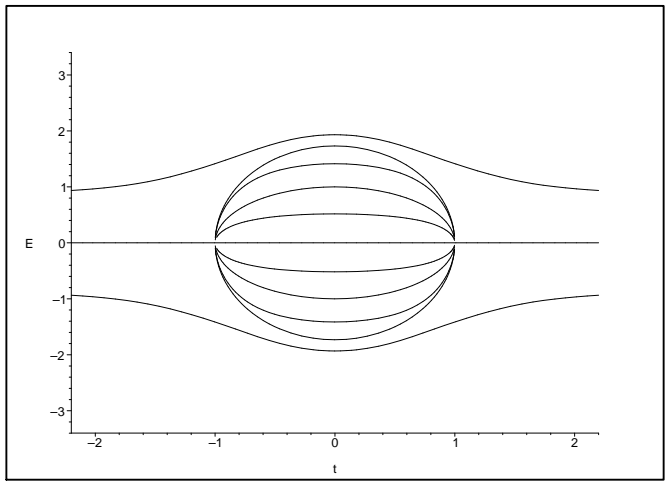,angle=270,width=0.5\textwidth}
\end{center}                         
\vspace{-2mm}\caption{Real eigenvalues of the $k=4$ Hamiltonian
$H^{(11)}(t,-t,t,-t)$ and the graphical localization of the
degenerate phase-transition EP points at $t=\pm 1$.
 \label{fi1}}
\end{figure}

Near the EP boundary $\partial {\cal D}$ some of the elements of the
diagonal metric $\Theta^{(diag)}$ of Proposition 1 will, for the
consistency reasons, almost vanish or almost diverge. We may expect,
therefore, that the most elementary parametric path of couplings
$\lambda=\mu=\ldots = t$ will certainly cross the EP boundary at
$t=\pm 1$. This expectation is confirmed by Fig.~\ref{fi1} in which
the $t-$dependence of the real spectrum as well as the
phase-transition-marking loss of its reality at $t=\pm 1$ are
displayed, for illustration, at $N=11$ and $k=4$.

The most interesting features illustrated by the latter picture seem
to be the $t-$independence of the exceptional level $E=0$, the
presence of the two outer ``spectator-like'' levels and, last but
not least, the nine-tuple EP degeneracy which occurs at $t=\pm 1$.
In a complementary step of analysis one can easily switch to
symbolic manipulations and derive the corresponding exact secular
equation
 \ben
 {{\it E}}^{11}- \left( 10-8\,{t}^{2} \right) {{\it E}}^{9}+ \left(
 36-58\,{t}^{2}+22\,{t}^{4} \right) {{\it E}}^{7}- \left(
 56-136\,{t}^ {2}+104\,{t}^{4}-24\,{t}^{6} \right) {{\it E}}^{5}+
 \een
\ben
 +\left( 35-114\,{ t}^{2}+132\,{t}^{4}-62\,{t}^{6}+9\,{t}^{8} \right)
 {{\it E}}^{3}-
 \left( 6-24\,{t}^{2}+36\,{t}^{4}-24\,{t}^{6}+6\,{t}^{8} \right)
  {\it E} =0\,.
   \een
On this ground we may confirm the above-mentioned graphical result
rigorously. Indeed, this secular equation degenerates to the trivial
relation ${{\it E}}^{11}- 2\, {{\it E}}^{9}=0$ at the two EP-marking
parameters $t=\pm 1$, etc.

\subsection{A typology of the unfoldings of the EP degeneracies}

In an attempt of exploring the small vicinity of the maximal EP
degeneracy at $\lambda^{(MEP)}=\mu^{(MEP)}=\nu^{(MEP)}
=\rho^{(MEP)}=1$ let us now fix one of the individual parameters
near the EP boundary $\partial {\cal D}$ and let us keep the
selected parameters, one by one, $t-$independent. This change will
define the four new phase-transition parametrizing paths of the
couplings. One may expect that the nine-fold degeneracy of the
energies of Fig.~\ref{fi1} at $t=\pm 1$ will unfold in different
ways forming the alternative phase-transition patterns.

\begin{figure}[h]                     
\begin{center}                         
\epsfig{file=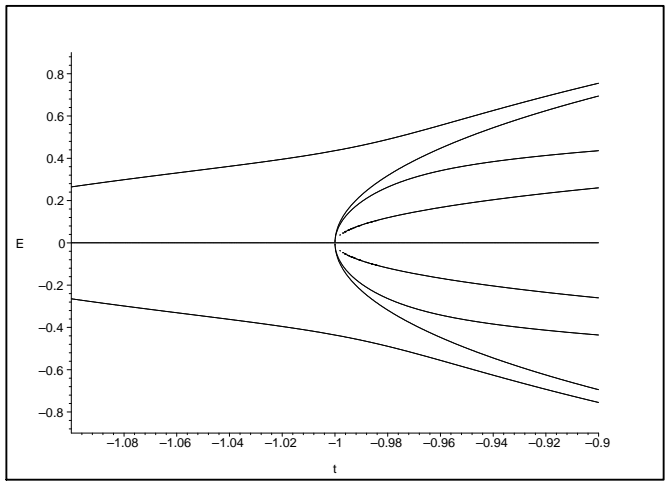,angle=270,width=0.5\textwidth}
\end{center}                         
\vspace{-2mm}\caption{The nine central real eigenvalues of the $k=4$
Hamiltonian $H^{(11)}(t,-t,t,-0.9)$ and the ``weakest'' unfolding of
the degeneracy near the phase-transition point $t=- 1$.
 \label{fig9a}}
\end{figure}

In Fig.~\ref{fig9a} we see the first $t-$dependent spectrum  in
which we fixed the innermost coupling $\rho=9/10$ and in which we
kept the other three couplings in the same form as above,
$\lambda=\mu=\nu= t$. With the two outer, ``spectator'' real levels
left out of the picture we see that the eight innermost levels
remain degenerate at $t=-1$ while the degeneracy of the remaining
pair gets shifted rather far to the left.

\begin{figure}[h]                     
\begin{center}                         
\epsfig{file=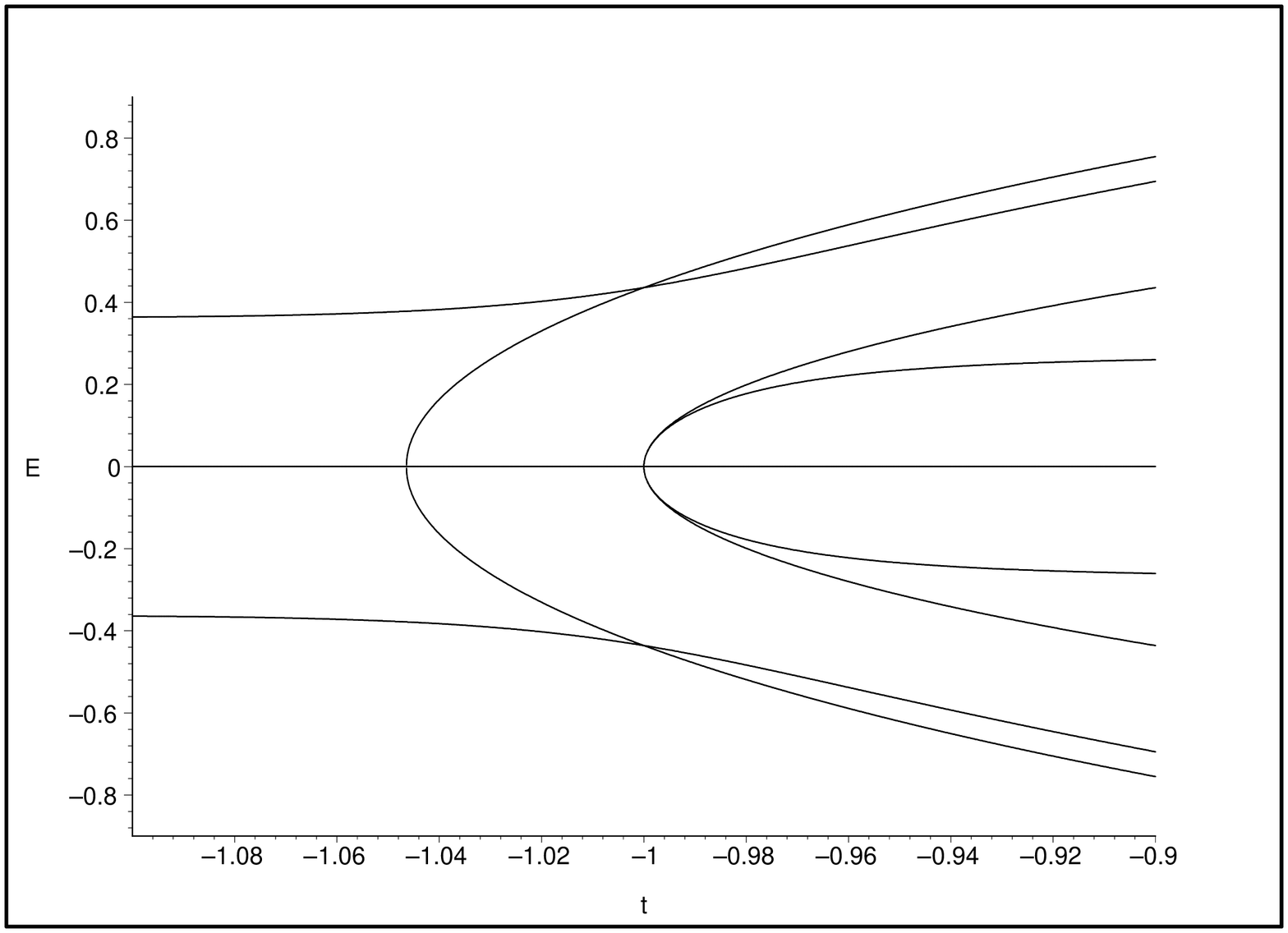,angle=270,width=0.5\textwidth}
\end{center}                         
\vspace{-2mm}\caption{The central real eigenvalues of the $k=4$
Hamiltonian $H^{(11)}(t,-t,0.9,-t)$ and the second form of the
unfolding of the degeneracy near the phase-transition point $t=- 1$.
 \label{fig9b}}
\end{figure}

In the subsequent picture provided by Fig.~\ref{fig9b} we see the
more thoroughly modified $t-$dependence of the spectrum which is
caused by the move to the next scenario in which we choose the
constant $\nu=9/10$ while keeping the remaining couplings variable
as above, $\lambda=\mu=\rho= t$. Ignoring now still the two
outermost spectator levels as less relevant, we observe that another
outer pair of the real levels has got separated at $t=-1$ and that
it only becomes degenerate and complexified more to the left.
Marginally, it is worth noticing that at the values of $t\ll -1 $
which already lie out of the picture (i.e., very far to the left)
the previously shifted second outer pair gets merged with the
respective upper or lower ``spectators'' so that merely the single,
exceptional constant energy $E=0$ remains real in the $|t|\gg 1$
asymptotic region.

\begin{figure}[h]                     
\begin{center}                         
\epsfig{file=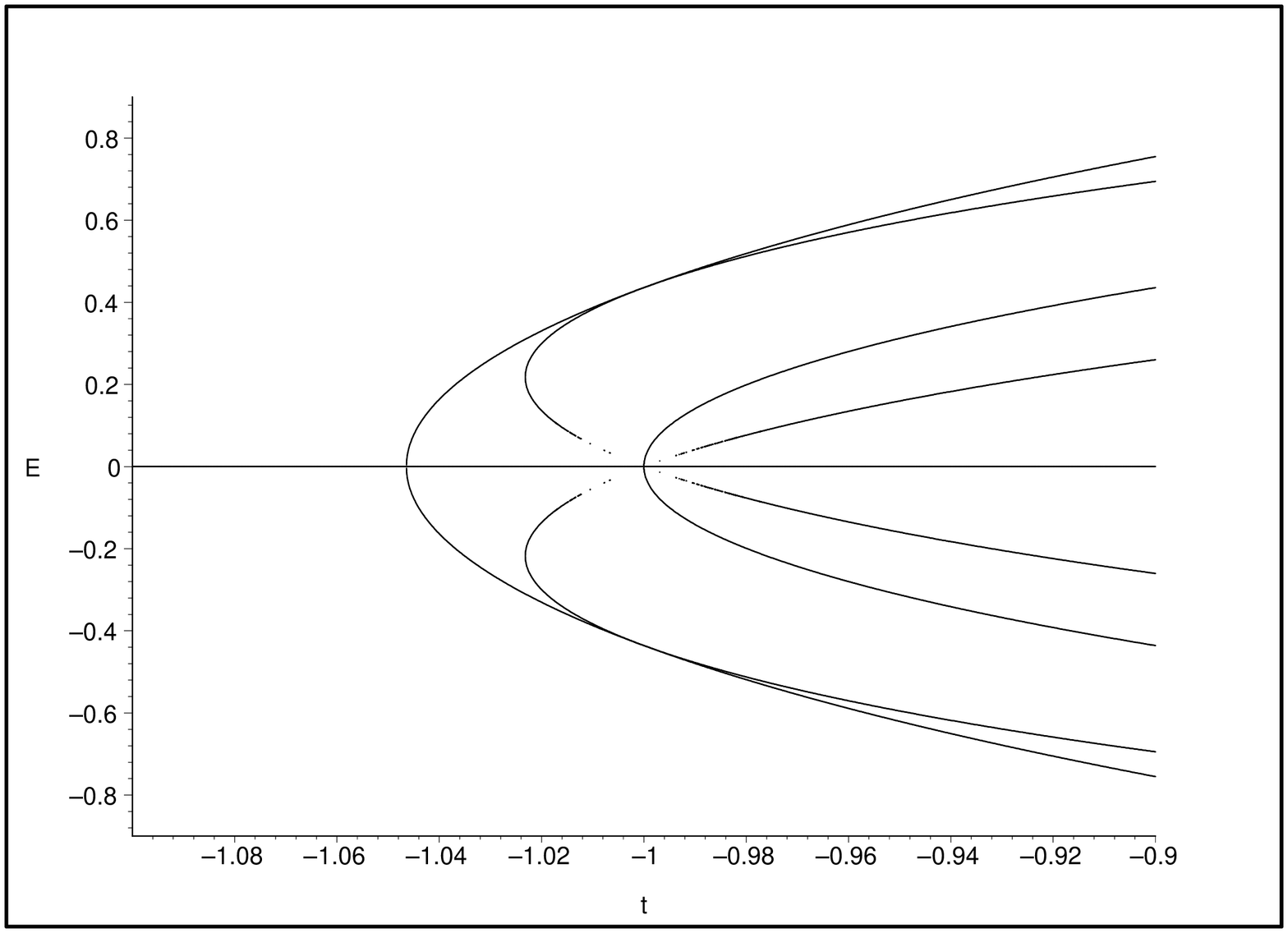,angle=270,width=0.5\textwidth}
\end{center}                         
\vspace{-2mm}\caption{The central real eigenvalues of the $k=4$
Hamiltonian $H^{(11)}(t,-0.9,t,-t)$ and the third form of the
unfolding of the degeneracy near the phase-transition point $t=- 1$.
 \label{fig9c}}
\end{figure}

A return to the asymptotic reality of the triplet of the energies
(including again two outermost spectators, i.e., not visible in
Fig.~\ref{fig9c}) is the phenomenon which characterizes, a bit
unexpectedly, the next choice of $\mu=9/10$ together with
$\lambda=\nu=\rho= t$. A graphical explanation is provided by the
parallel change of behaviour of the levels near the MEP degeneracy.
In Fig.~\ref{fig9c} we see that the sequential unfolding of this
degeneracy further continues in a way which is a bit more subtle. In
fact, the ``expectable'' complexification of the further two
nontrivial innermost energy trajectories  gets replaced by their
mere crossing, followed by the two separate subsequent
complexifications which only occur again a bit later, i.e., further
to the left.

%

\begin{figure}[h]                     
\begin{center}                         
\epsfig{file=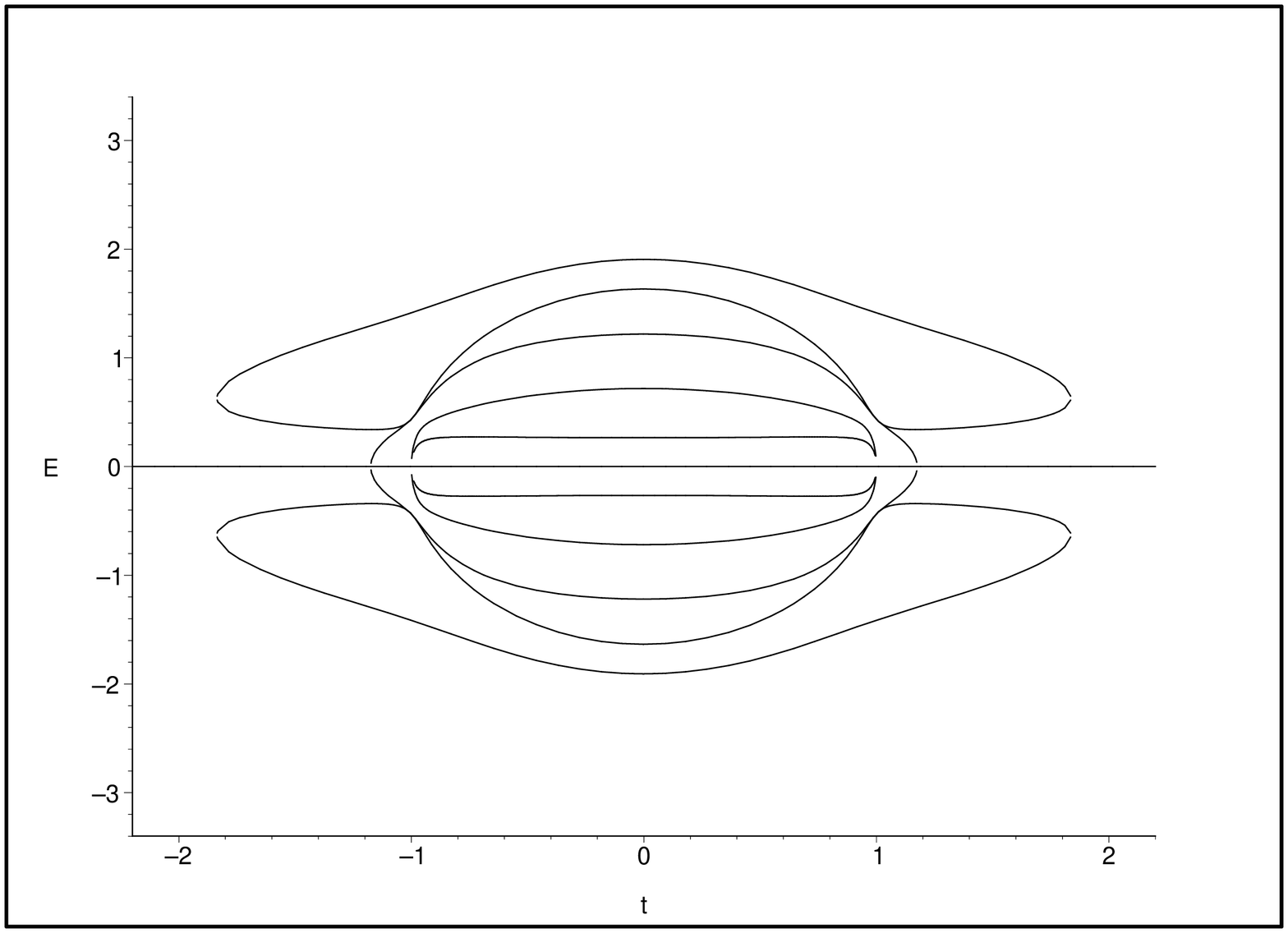,angle=270,width=0.5\textwidth}
\end{center}                         
\vspace{-2mm}\caption{Real eigenvalues of the $k=4$ Hamiltonian
$H^{(11)}(0.9,-t,t,-t)$ and the strongest form of the unfolding of
the MEP degeneracy at the phase-transition points $t=\pm 1$.
 \label{fig9}}
\end{figure}

After the last possible choice of $\lambda=9/10$ with $\mu=\nu=\rho=
t$ all the nontrivial levels get again complex at the sufficiently
large $|t|\gg 1$. The whole real spectrum is shown in
Fig.~\ref{fig9} where we see that the complexification now involves
the sextuplet of  outer energy levels. The extreme phase transition
pattern of Fig.~\ref{fi1} is now  changed most thoroughly. The
original degeneracy of the spectrum survives just in the weakest
form at $t=\pm 1$. This and similar observations may be also deduced
from the $t=\pm 1$ form of the secular equation,
 \ben
{{\it E}}^{11}-{\frac {119}{50}}\,{{\it E}}^{9}+{\frac {7961}{10000}
}\,{{\it E}}^{7}-{\frac {361}{5000}}\,{{\it E}}^{5}=0\,.
 \een
Due to the unexpected exact solvability of this $t=\pm 1$ algebraic
equation one can confirm the expected presence of the five
degenerate inner roots $E=0$ and also the less expectable double
degeneracy of the two other, non-vanishing roots $E=\pm
\sqrt{19}/10$ which moved away from zero. The last though, possibly,
just marginal surprise is that the last two non-degenerate, outer
energy roots $E=\pm \sqrt{2}$ did not move after the change of path
at all.

\section{Summary}

In an overall applied-mathematics context and, more explicitly,
within the framework of the use of the crypto-Hermitian
representations of observables in quantum mechanics \cite{SIGMA} our
present paper and main model-building message may be read as based
on the following three methodical assumptions, viz.,

\begin{itemize}

\item
{\{1\}} the requirement of the feasibility of constructive
considerations

\item
{\{2\}} a correspondence-principle connection

\item
{\{3\}} an offer of insight.

\end{itemize}

 \noindent
We interpreted point \{1\} as our convenient restriction of the
Hilbert spaces in question to the finite-dimensional ones. In
section \ref{newmodel} we realized the second assumption \{2\} via a
``derivation'' of our main difference-operator $N < \infty$
toy-model example from its differential-operator $N=\infty$
predecessor of Ref.~\cite{david} (cf. also Ref.~\cite{complete} for
more details). Thirdly, in connection with item \{3\} it is worth
emphasizing that our present results just reconfirmed that the
symbolic manipulations and the related advanced software (like
MAPLE, etc) became, with time, an inseparable {\em condition sine
qua non} of all the similar, manifestly constructive (i.e.,
basically, applied-linear-algebra) projects.

Our present concrete results were aimed, basically, at a deeper
understanding of the spectral features of Hamiltonians sampled by
the new model $H^{(N)}({\lambda, -\mu,\ldots})$ of section
\ref{newmodel}. The motivation of our study was firmly rooted in the
underlying physics. Briefly, it was aimed at a constructive analysis
of the parametric domain ${\cal D}$ of the unitarity of the
underlying hypothetical physical quantum system.

Such a global purpose and aim have been achieved in several
directions. Firstly, in contrast to the most common and plainly
Hermitian toy models (where, typically, ${\cal D}\ \equiv \
\mathbb{R}^d$ has no accessible boundary) we (re-)emphasized that
the domains ${\cal D}$ of our present interest were, typically,
compact, possessing a nontrivial EP boundary {\em alias} horizon
$\partial {\cal D} \neq \emptyset$. Secondly, we demonstrated that
the constructive study of many features of these horizons may be
rendered feasible via an interactive use of a suitable graphical
software, of a sufficiently advanced computer arithmetics and, first
of all, of certain extensive symbolic manipulation package.

Speaking in technical terms we used MAPLE and profited from its
Gr\"{o}bner basis facilities, etc. Thus, although our original
motivation came from the physical background (concerning, typically,
the questions of the stability of quantum systems), our concrete
main tasks (typically, the manipulations with secular polynomials)
and results (typically, the recurrent construction of non-Dirac
metrics $\Theta\neq I$, etc) were of a more mathematical nature.

Within this framework, a subsequent return to physics might be
inspired, in the nearest future, by the ``next-step'' transition to
the complex-matrix model-building. Such a next-step enrichment of
the tunable dynamics would already lead us very close to
experimental setups. Typically, they might be explored, using
coupled waveguides, in a way sampled, more concretely, in
Ref.~\cite{fortinbras}.

In conclusion let us mention that in the  future study of the model
the emphasis may be expected to get shifted, formally speaking,
beyond the horizons  $\partial {\cal D}$ and out of the
quantum-stability domains ${\cal D}$. The emergent complex spectra
and unstable physical mechanisms  seem to open new challenging
questions. Curiously enough, we are witnessing an enormous increase
of interest in similar phenomena not only in the theoretical studies
of quantum catastrophes \cite{catast} and in many innovative and
non-standard practical quantum-model analyses
\cite{Nimrod,Hubb,Ingrid,Longhi} but even far beyond the quantum
physics itself and, in particular, as we already mentioned, in
classical optics \cite{Makris,Makrisbe}.


\subsection*{Acknowledgments}

Research supported by the GA\v{C}R grant Nr. P203/11/1433.


\end{document}